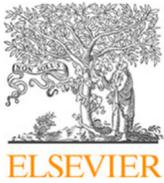
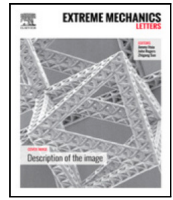

# HyperCAN: Hypernetwork-driven deep parameterized constitutive models for metamaterials

Li Zheng [a], Dennis M. Kochmann [a], Siddhant Kumar [b],*

[a] *Mechanics & Materials Laboratory, Department of Mechanical and Process Engineering, ETH Zürich, 8092, Switzerland*
[b] *Department of Materials Science and Engineering, Delft University of Technology, 2628 CD Delft, Netherlands*

A B S T R A C T

We introduce HyperCAN, a machine learning framework that utilizes hypernetworks to construct adaptable constitutive artificial neural networks for a wide range of beam-based metamaterials exhibiting diverse mechanical behavior under finite deformations. HyperCAN integrates an input convex neural network that models the nonlinear stress–strain map of a truss lattice, while ensuring adherence to fundamental mechanics principles, along with a hypernetwork that dynamically adjusts the parameters of the convex network as a function of the lattice topology and geometry. This unified framework demonstrates robust generalization in predicting the mechanical behavior of previously unseen metamaterial designs and loading scenarios well beyond the training domain. We show how HyperCAN can be integrated into multiscale simulations to accurately capture the highly nonlinear responses of large-scale truss metamaterials, closely matching fully resolved simulations while significantly reducing computational costs. This offers new efficient opportunities for the multiscale design and optimization of truss metamaterials.

## 1. Introduction

Architected materials have unlocked immense design possibilities by manipulating their architectures across scales. With recent advances in additive manufacturing, mechanical metamaterials, featuring beam- [1,2], plate- [3], or shell-based [4–6] architectures, have revolutionized material design by enabling tailored properties across diverse applications, including ultralight structures [7,8], auxetics [9–11], impact protective systems [12,13], and acoustic waveguides [14,15]. Periodic beam-based metamaterials, commonly referred to as truss metamaterials or truss lattices and characterized by periodic arrangements of interconnected beam networks, offer a vast design landscape for previously unattainable properties and functionality by exploiting the topological and geometrical attributes of small-scale architectures [16–20]. To fully exploit the design potential of truss metamaterials, particularly in applications such as biomimetic implants [21,22], soft robotics [23,24], and energy absorption [25,26], it is crucial to understand their mechanical behavior under large deformation. A major challenge in modeling such truss lattices is the intensive computations required to capture the responses of individual structural members in the unit cell due to the complexity arising from geometric and material nonlinearity and associated instabilities [27,28]. This calls for effective constitutive models which, depending on the underlying unit cell design, facilitate efficient multiscale simulation and optimization.

Traditional modeling approaches have relied on formulating effective constitutive relations based on intuitive understanding and physical assumptions [29–33], with material-dependent parameters calibrated through phenomenological observations or experimental tests. Consequently, phenomenological constitutive models are often constrained by available experimental data and a-priori assumptions, lacking the flexibility to adapt to new materials or untested complex scenarios. Over the past decade, data-driven methods have emerged as a powerful and versatile alternative to circumvent the need for explicit formulations of constitutive relations [34]. By directly harnessing datasets of strain energies or stress responses, data-driven constitutive models approximate the constitutive behavior by, e.g., polynomial [35] or spline interpolation [36,37], Gaussian process regression [38,39], or reduced-order models [40,41]. Notably, artificial neural networks (NNs) have demonstrated remarkable success in learning constitutive laws across various materials for a range of material behaviors, from small strain elasticity [42] and finite strain hyperelasticity [43–45], to plasticity [46,47] and viscoelasticity [48], demonstrating exceptional capabilities as universal approximators. Despite the efficiency of NNs in capturing intricate nonlinear material responses, unrestricted NNs may easily violate underlying mechanical constraints and therefore lack physical interpretability. To overcome this limitation, there is a growing interest in integrating ML tools with physical knowledge to build constitutive models that are both data-efficient and well-grounded in physical principles [49–54].

* Corresponding author.
  *E-mail addresses:* dmk@ethz.ch (D.M. Kochmann), sid.kumar@tudelft.nl (S. Kumar).

Strategies for developing physics-informed NN constitutive models for elastic finite strains typically involve learning the hyperelastic strain energy density function $W$ with strain measures as input, such as the deformation gradient $\boldsymbol{F}$ or the Green–Lagrange strain tensor $\boldsymbol{E}$ (equivalently, the right Cauchy–Green deformation tensor $\boldsymbol{C}$). Their respective conjugate stress measures, the first Piola–Kirchhoff stress $\boldsymbol{P}$ and the second Piola–Kirchhoff stress $\boldsymbol{S}$, can be derived by differentiating the strain energy density. To ensure the NN model's adherence to fundamental physical considerations while maintaining flexibility and generality, physics-informed inductive biases in selecting the parameterization and model architecture can be advantageous. For instance, by expressing the energy function using $\boldsymbol{E}$ or $\boldsymbol{C}$ as inputs, the model inherently satisfies the objectivity constraints of $W$ per construction [50,55,56]. Furthermore, the ellipticity of W in $\boldsymbol{F}$ is an important consideration, as it ensures material stability [57]. In general, directly enforcing a model to be elliptic a priori is cumbersome; hence, a common approach is to instead enforce polyconvexity, which implies ellipticity while not being too strong to exclude the non-uniqueness of the energy minimizer [58,59]. Several data-driven constitutive models based on *input convex neural networks* (ICNNs) [60] have been proposed to construct polyconvex hyperelastic potentials [6,45,61,62]. By incorporating polyconvex invariants (or the deformation gradient along with its cofactor and its determinant) as inputs to the neural network, ICNNs automatically satisfy the polyconvexity constraint through their architectural design and selection of activation functions. Polyconvex NNs have proven successful in accurately predicting anisotropic responses at finite deformations for various materials and, more importantly, extrapolating to untested deformation states, demonstrating that leveraging physical principles significantly enhances the model's interpretability and generalization capabilities.

Despite their efficiency and reliability, most existing NN constitutive models have been restricted to a single type of material. Once trained on responses for one specific material (or metamaterial unit cell), the model is fixed and cannot be directly employed for different materials without retraining. This is highly inefficient when considering the tremendous design space of metamaterials and their intricate structure-to-property maps. One potential solution is to incorporate constitutive models with an explicit parameterization [60], that is, the parameterized model takes both the parameters $\boldsymbol{p}$ characterizing the material and the deformation gradient $\boldsymbol{F}$ as inputs and predicts the strain energy density function as $W = W(\boldsymbol{F}, \boldsymbol{p})$. The parameterized constitutive NN allows the model to account for variations in topological [6,49] or material properties [63] when predicting material responses, thereby broadening its applicability across different material configurations.

Constitutive modeling for truss metamaterials, however, presents unique challenges and has remained largely under-explored. Truss lattices provide an extensive and diverse design space, enabled by variations in lattice topologies (i.e., connectivity of the beam network) and geometrical features (i.e., node placements, strut cross-sections, etc.). Despite this vast potential, existing works are often confined to a small catalog of lattices [61,64,65] due to the lack of generalized parameterization of trusses, not fully leveraging the available design freedom. Additionally, beam lattices exhibit complex nonlinear mechanical behavior at large strains, primarily due to their geometrical nonlinearities and mechanical instabilities such as beam buckling. The latter is highly sensitive to the configurations; e.g., small changes in the unit cell topology or strut orientation can drastically shift the deformation mechanisms (stretch- vs. bending-dominated) of truss lattices [17,66]. Such sensitivity requires the NN model to effectively learn not only the individual effects of structural features and deformation states but also their combined effects on mechanical behaviors. This becomes particularly challenging for beam networks due to their often high-dimensional and discrete parameterization. More recently, graph representations have become prominent as efficient tools for capturing the 3D structural features of truss lattices [19,67,68], used, e.g., in conjunction with graph neural networks to predict the mechanical response of those structures. Nevertheless, existing studies predominantly address only linear and small-strain responses [69,70], or nonlinear responses subject to one specific loading scenario such as uniaxial compression/tension [71,72], thus limiting their applicability to different loading conditions. To the best of our knowledge, there exists no constitutive modeling approach that is adaptable for various truss configurations across different topological and geometrical features as well as for different loading conditions, while fulfilling key constitutive conditions.

To address the aforementioned challenges, we propose a machine learning (ML) framework for constructing effective hyperelastic constitutive models of beam-based metamaterials at finite strains, covering a broad spectrum of truss architectures and anisotropic material behavior. Distinct from existing works that directly train a static NN constitutive model with stress responses for specific lattice configurations, the goal of our generalized constitutive modeling framework is to develop adaptable constitutive models capable of predicting the challenging behavior of truss metamaterials undergoing large deformation and instabilities, as well as generalizing and extrapolating across a vast range of metamaterial architectures and loading conditions beyond the training domain. To this end, we propose *HyperCAN*, which utilizes hypernetworks [73] to learn <u>c</u>onstitutive <u>a</u>rtificial <u>n</u>eural networks for various metamaterial architectures by leveraging previously acquired knowledge. Hypernetworks have shown promising results in developing NNs with greater flexibility, adaptability, and generalization in various deep learning tasks, including multitasking [74,75], transfer learning [76], and domain adaptation [77,78]. By employing hypernetworks to customize weights of a diverse family of constitutive NNs, our approach provides a unified framework for learning constitutive models that can efficiently adapt to unseen scenarios, bypassing the need for fine-tuning multiple constitutive models.

Our proposed HyperCAN framework consists of two main components: a hypernetwork and a constitutive neural network. The constitutive NN uses an ICNN core, which encodes important mechanical considerations to construct the hyperelastic energy potential. The architectural design of ICNNs and proper parameterization automatically guarantee thermodynamic consistency, material objectivity, a stress-free reference configuration, and local material stability. The hypernetwork takes as input a unit cell design and aims to generate as output the weights of a constitutive NN. Both networks are trained end-to-end through backpropagation, which enables the hypernetwork to capture the intricate dependencies across various truss lattices and to adjust the constitutive relations accordingly. We demonstrate that our HyperCAN approach is not only capable of accurately describing the nonlinear hyperelastic effective behavior of various truss lattices but also more flexible and adaptive, showcasing great generalization and robustness to **unseen unit cell configurations and unseen loading scenarios**. Finally, the NN constitutive model is integrated into a multiscale finite element simulation for truss metamaterials, effectively capturing the highly nonlinear macroscopic responses of truss lattices with significantly reduced computational efforts compared to fully resolved simulations. Altogether, our proposed generalized ML framework opens new avenues for constructing efficient, highly adaptable, and physically sensible data-driven constitutive models for a broad range of architected materials. This, combined with the tremendous design space of the chosen graph-based truss representation, paves the way for fully exploring the potential of the multiscale design of metamaterials.

## 2. Methods

### 2.1. Constitutive requirements for hyperelasticity

To ensure that the NN constitutive model satisfies the necessary physical constraints, let us briefly discuss the theoretical basis of hyperelastic constitutive modeling relevant to this work. A hyperelastic constitutive model describes the material behavior up to large strains and can be characterized by a strain energy density $W = W(F)$ stored within a body $\Omega \subset \mathbb{R}^3$, where $F \in \mathrm{GL}_+(3)$ denotes the deformation gradient. The first Piola–Kirchhoff stress $P(F)$ is derived as

$$P(F) = \frac{\partial W}{\partial F}(F). \tag{1}$$

To construct a physically valid hyperelastic strain energy density, the following constitutive requirements must be fulfilled:

- In the absence of deformation (i.e., in the reference configuration where $F = I$), the stress should vanish, i.e.,

$$P(I) = 0. \tag{2}$$

- The strain energy density must be objective, i.e., $W$ must be invariant under any rotation tensor $Q$ (satisfying $Q^T Q = I$ and $\det Q = 1$) of the deformation gradient $F$:

$$W(QF) = W(F) \quad \forall \ F \in \mathrm{GL}_+(3), \ Q \in \mathrm{SO}(3). \tag{3}$$

- The ellipticity (also known as rank-one convexity) condition of the strain energy density

$$(\xi \otimes \eta) : \frac{\partial^2 W(F)}{\partial F^2} : (\xi \otimes \eta) \geq 0 \quad \forall \ \xi, \eta \in \mathbb{R}^3 \tag{4}$$

ensures pointwise material stability. However, enforcing ellipticity a priori in model formulations is challenging. Consequently, a common approach in constitutive modeling is to instead enforce polyconvexity, which implies ellipticity. Specifically, $W(F)$ is polyconvex if and only if there exists a function $\mathcal{P} : \mathrm{GL}_+(3) \times \mathrm{GL}_+(3) \times \mathbb{R} \to \mathbb{R}$ such that

$$W(F) = \mathcal{P}(F, \mathrm{Cof} \ F, \det F). \tag{5}$$

Based on Eq. (5), a straightforward approach to enforce polyconvexity is to formulate $W$ as convex functions of input $\{F, \mathrm{Cof} \ F, \det F\}$. This, however, violates the material objectivity in Eq. (3), since the deformation gradient and its cofactor are not invariant to the orientation of an observer. As a remedy, data augmentation approaches that introduce randomly sampled observers can be applied to approximate the objectivity condition [61,64]. An alternative is to employ polyconvex and objective invariants as inputs [45,49,61,79], which naturally fulfills the objectivity condition. However, the model capacity strongly depends on the predefined choice of invariants and specific material symmetry groups, which poses challenges in generalizing to a wide range of anisotropic materials or unseen loading scenarios beyond the training domain.

Therefore, considering that objectivity is an essential property that must be fulfilled non-trivially, we choose to express the strain energy density as a function of the Green–Lagrange strain tensor $E = (F^T F - I)/2$, i.e., in the form of $W = W(E)$, which inherently satisfies the material objectivity condition. Furthermore, the convexity of $W$ in $E$ implies the local convexity of $W$ in $F$ (see Refs. [62,80] for proofs). This is preferable over the global convexity of $W$ in $F$, as the latter excludes the non-uniqueness of solutions due to structural buckling. These considerations motivate the use of ICNNs to construct a constitutive model that outputs a convex strain energy density $W_\theta^{\mathrm{NN}}(E)$ as a function of the Green–Lagrange strain tensor $E$. As such a NN constitutive model may still violate the stress-free reference configuration condition in Eq. (2) without explicit constraints, we include linear correction terms [45,62] in the strain energy density, ensuring that the model exhibits zero stress and zero strain energy density in the absence of deformation by construction. In summary, we propose to formulate the strain energy density as

$$W(E) = W_\theta^{\mathrm{NN}}(E) + H : E + W^0, \tag{6}$$

where $W_\theta^{\mathrm{NN}}(E)$ represents the ICNN model parameterized by $\theta$. The energy correction term $W^0$ is a constant scalar that ensures zero strain energy density at zero deformation $F = I$, or equivalently, $E = 0$, i.e.,

$$W^0 = -W_\theta^{\mathrm{NN}}(E)\Big|_{E=0} \quad \to \quad W(0) = 0. \tag{7}$$

The second Piola–Kirchhoff stress tensor and the components of the tangent modulus tensor can be derived by differentiating the strain energy density as

$$\begin{aligned} S(E) &= \frac{\partial W}{\partial E}(E) = \frac{\partial W_\theta^{\mathrm{NN}}}{\partial E}(E) + H, \\ \mathbb{C}(E) &= \frac{\partial S}{\partial E}(E) = \frac{\partial^2 W_\theta^{\mathrm{NN}}}{\partial E \partial E}(E) \quad \text{with} \quad \mathbb{C}_{ijkl} = \frac{\partial^2 W_\theta^{\mathrm{NN}}}{\partial E_{ij} \partial E_{kl}}, \end{aligned} \tag{8}$$

The stress correction term involving second-order tensor $H$ is chosen such that the model is stress-free for $F = I$ (equivalently, $E = 0$), i.e.,

$$H = -\frac{\partial W_\theta^{\mathrm{NN}}}{\partial E}\Big|_{E=0} \quad \to \quad S(0) = 0. \tag{9}$$

Therefore, the neural network hyperelastic constitutive model defined in Eq. (6) ensures thermodynamic consistency, preserves the local convexity of the strain energy density, and fulfills both the stress-free reference configuration condition and the objectivity condition—without over-constraining the strain energy density to admit instability (as expected, e.g., from the effective response of truss metamaterials undergoing buckling).

### 2.2. ICNN-based constitutive modeling

We here briefly introduce the ICNN [60] architecture, which forms the core of our NN-based constitutive model. The architectural design and activation functions of the ICNN are specially chosen to ensure that it outputs a strain energy density $W_\theta^{\mathrm{NN}}(E)$ that exhibits local convexity, thus encoding the required physical knowledge into the constitutive model (for proofs, see Ref. [60]). Here, we consider a fully convex NN based on a multilayer perceptron (MLP) with $k$ hidden layers. Given the deformation gradient $F$, the model first computes the Green–Lagrange strain tensor $E = (F^T F - I)/2$, which is then vectorized as $y = [E_{11}, E_{12}, E_{13}, E_{21}, E_{22}, E_{23}, E_{31}, E_{32}, E_{33}]^T \in \mathbb{R}^{d_0}$ (with $d_0 = 9$) and passed to the ICNN core. The $i$th hidden layer ($i = 1, 2, \ldots, k$) of size $d_i$ takes the output from the previous layer, $z^{(i-1)}$, and computes the output as

$$z^{(i)} = \phi\big(\underbrace{\sigma(\mathcal{W}_z^{(i)}) z^{(i-1)}}_{\text{convex layers}} + \underbrace{\mathcal{W}_y^{(i)} y}_{\text{passthrough layers}} + \underbrace{b^{(i)}}_{\text{bias}}\big) \tag{10}$$

where $\mathcal{W}_z^{(i)} \in \mathbb{R}^{d_i \times d_{i-1}}$ and $b^{(i)} \in \mathbb{R}^{d_i}$ are the weight matrix and the bias vector of fully connected layers, respectively, $\mathcal{W}_y^{(i)} \in \mathbb{R}^{d_i \times d_0}$ is the weight matrix of the passthrough layers that directly connect the inputs $y \in \mathbb{R}^{d_0}$ to the deeper hidden layers, and $\phi$ and $\sigma$ are nonlinear element-wise functions.

The above network ensures convexity through several key restrictions. First, all weights in the fully connected layers are constrained to be *non-negative*. This is achieved by applying a nonlinear non-negative function $\sigma$ to the weight matrix $\mathcal{W}_z^{(i)}$, while allowing $\mathcal{W}_y^{(i)}$ to be negative, thereby maintaining the model's predictive power without compromising convexity. Second, all activation functions $\phi$ must be *convex and non-decreasing*. In addition, $W_\theta^{\mathrm{NN}}$ needs to be at least twice differentiable to enable the calculation of the tangent modulus $\mathbb{C}$ as shown in Eq. (8), which requires activation functions to have non-vanishing second derivatives. The aforementioned conditions motivate the choice of squared-softplus and softplus activation functions for $\phi$ and $\sigma$, respectively:

$$\phi(x) = \alpha \left(\log(1 + e^x)\right)^2 \quad \text{and} \quad \sigma(x) = \beta \log(1 + e^x), \tag{11}$$

where $\alpha, \beta \in \mathbb{R}^+$ are hyperparameters that control the curvature of functions $\phi(x)$ and $\sigma(x)$, respectively. The squaring of the softplus function $\phi$ provides smooth, non-negative, and convex activation, ensures stable training by preventing vanishing gradients [81–83], and enhances the expressiveness and accuracy of the constitutive model, e.g., by avoiding spurious local oscillations when predicting stress responses [45,62].

In summary, we construct the constitutive model $W_\theta^{\text{NN}} = \mathcal{F}(E; \theta)$ based on ICNNs, where $\theta = \{\mathcal{W}_z^{(i)}, \mathcal{W}_y^{(i)}, \boldsymbol{b}^{(i)} : i = 1, \ldots, k\}$ represents the set of unknown NN weights and biases. The strain energy density $W_\theta^{\text{NN}}$ relies solely on the objective strain measure $E$, thus inherently satisfying the objectivity condition in Eq. (3). By applying a non-negative transformation function $\sigma$ to weights and employing convex non-decreasing activation function $\phi$, the NN constitutive model preserves the local convexity of the strain energy density, thus ensuring local material stability. Additionally, passthrough connections are integrated into each hidden layer to address the limited expressiveness due to the non-negativity constraint of layer weights. The resulting mechanics-informed NN model seeks to capture the complex, intractable mechanical relations and to accurately predict the highly nonlinear material behavior at finite strains. The predictive capabilities and physical insights provided by the NN constitutive model can further be exploited to enhance the stability and efficiency of nonlinear multiscale simulations of architected materials.

### 2.3. Generation of the metamaterial design space with diverse nonlinear responses

To develop efficient NN-based constitutive models for various truss metamaterials, we aim to construct a design space that encompasses a wide array of truss configurations, exhibiting diverse mechanical responses. We particularly focus on beam lattices based on the 3D periodic tessellation of a cubic representative volume element (RVE). We modify the graph-based representation of truss lattices, following Ref. [19], which allows for a unified and systematic representation of periodic trusses. By assuming symmetries across three mutually orthogonal planes, the RVE is partitioned into eight equal octants. Within each octant, we define a total of 14 possible node placements: 8 vertex nodes $\{v_0, v_1, v_2, v_3, v_4, v_5, v_6, v_7\}$ and 6 face nodes $\{f_0, f_1, f_2, f_3, f_4, f_5\}$, as shown in Fig. 1a. The vertex nodes are fixed to maintain connectivity on the outer boundaries of the unit cell, whereas the face nodes can freely move within the face. Each truss can be naturally translated into a graph representation $G = (\boldsymbol{A}, \boldsymbol{x})$, where $\boldsymbol{A} \in \{0,1\}^{14 \times 14}$ denotes the adjacency matrix representing all beam connections and $\boldsymbol{x} \in \mathbb{R}^{14 \times 3}$ denotes the node positions. Beginning with 6 elementary unit cells as initial designs, we apply an iterative stochastic perturbation algorithm to generate new lattice designs by randomly inserting or removing nodes and connections. By perturbing both the truss connectivity and node positions, we construct a database of truss lattices containing 6,000 unique structures, achieving a reasonable balance between computational efficiency and the diversity of training space. Details regarding the parameterization and creation of the truss design space can be found in Ref. [19]. Fig. 1b shows representative examples of truss structures in our dataset.

To characterize the effective nonlinear responses of truss lattices, we perform finite element (FE) homogenization with periodic boundary conditions, leveraging the open-source FE code *ae108* [84]. Each individual strut is modeled as a linear elastic corotational (Euler–Bernoulli) beam element with a constant circular cross-section and Young's modulus $E_s = 1$ (undergoing large rotations but small strains). The beam thickness of all struts is varied to maintain a constant relative density $\rho = 0.025$ across all structures. Effective deformation gradients $\boldsymbol{F}$ are applied on average onto a unit cell through displacement-based periodic boundary conditions for $T > 0$ quasistatic load steps at uniform strain increments. For each load step, we compute the effective strain energy density $W$ by averaging the strain energy stored in the beams over the unit cell. The first and second Piola–Kirchhoff stress tensors can be derived from the reaction forces on the outer nodes of the unit cells. Fig. 1b shows the homogenized stress–strain responses of representative unit cells subject to uniaxial compression, exhibiting distinct nonlinear behaviors, which underscores the challenges in learning the highly complex structure-dependent deformation mechanisms. Unlike existing approaches [72,85] whose NNs serve as surrogate models to predict stress responses for specific loading paths (e.g., uni-/biaxial compression), we aim to develop generalized constitutive models capable of handling various combinations of structural configurations and loading conditions. In this study, we focus on the compressive and shear responses of truss structures, showcasing critical phenomena such as buckling instabilities and localized deformation. These characteristics are essential for assessing the constitutive model's robustness and accuracy in predicting highly nonlinear and anisotropic mechanical behaviors of truss structures. Additionally, the constitutive model adopts a generalized formulation, enabling straightforward extension to other deformation modes. To facilitate training the constitutive model, for each structure, we consider the following deformation paths: uniaxial compression (UC), biaxial compression (BC), and simple shear (SS). The corresponding deformation gradients are given by (without summation convention)

$$\begin{aligned}
\mathbf{F}^{\text{UC}}(\lambda) &= \boldsymbol{I} + \lambda \boldsymbol{e}_i \otimes \boldsymbol{e}_i, & \text{with} \quad i &\in \{1,2,3\}, \quad \lambda \in [-0.25, 0]; \\
\mathbf{F}^{\text{BC}}(\lambda) &= \boldsymbol{I} + \lambda_1 \boldsymbol{e}_i \otimes \boldsymbol{e}_i + \lambda_2 \boldsymbol{e}_j \otimes \boldsymbol{e}_j, & \text{with} \quad i,j &\in \{1,2,3\}, j \neq i, \quad \lambda_1, \lambda_2 \in [-0.25, 0]; \\
\mathbf{F}^{\text{SS}}(\lambda) &= \boldsymbol{I} + \lambda \boldsymbol{e}_1 \otimes \boldsymbol{e}_2, & \text{with} \quad \lambda &\in [0, 0.5],
\end{aligned} \tag{12}$$

where $\lambda$ is a loading parameter uniformly sampled from the given ranges. Furthermore, to assess the generalization capabilities of the proposed constitutive modeling framework, we consider an unseen validation loading scenario, which involves a more complex deformation path: triaxial compression (TC), characterized by the following deformation gradient

$$\mathbf{F}^{\text{TC}}(\lambda) = \boldsymbol{I} + \lambda_1 \boldsymbol{e}_1 \otimes \boldsymbol{e}_1 + \lambda_2 \boldsymbol{e}_2 \otimes \boldsymbol{e}_2 + \lambda_3 \boldsymbol{e}_3 \otimes \boldsymbol{e}_3, \quad \text{with} \quad \lambda_1, \lambda_2, \lambda_3 \in [-0.25, 0]. \tag{13}$$

The validation path TC remains unseen by the model and does not contribute to the training process. For each design, we consider a total of $K$ different loading paths, each discretized into $T$ load steps. For each path, we simulate the effective response of unit cells and collect the resulting $T$ values of the hyperelastic strain energy density and second Piola–Kirchhoff stress $\{(G, E_t^{(k)}, S_t^{(k)}, W_t^{(k)}) : k = 1, \ldots, K; t = 1, \ldots, T\}$ to construct a comprehensive structure-response dataset.

### 2.4. Learning parameterized constitutive models for truss metamaterials via hypernetworks

To address the complexities involved in constructing constitutive models for a broad range of truss metamaterials, we propose the HyperCAN framework, which employs a hypernetwork [73] to learn the abstract parameter space characterizing the constitutive models. Specifically, hypernetworks utilize a meta-learning strategy, which decomposes the task of learning constitutive models for various material configurations into two separate networks: a hypernetwork and a target/main network. The target network $\mathcal{F}$ is designed to perform the actual learning tasks, such as classification and regression, whereas the hypernetwork $\mathcal{H}$ dynamically generates the weights for the target network $\mathcal{F}$ based on input conditions. Distinct from existing works [56,61] on NN-based constitutive modeling that directly learn and update the constitutive NN parameters $\theta$ for a specific material type during training, we propose to use the hypernetwork to predict each individual set of $\theta$, including weights and biases for layers in $W^{NN}_\theta$, taking the input material features that characterize, e.g., the material properties or the microstructure geometries. This reparameterization allows the hypernetwork to capture the complex parameter space for different metamaterials with a significantly reduced number of parameters [86,87], leading to better adaptability and generalization of the target network to new, unseen data, as the hypernetwork can dynamically adjust its output weights to varying input conditions.

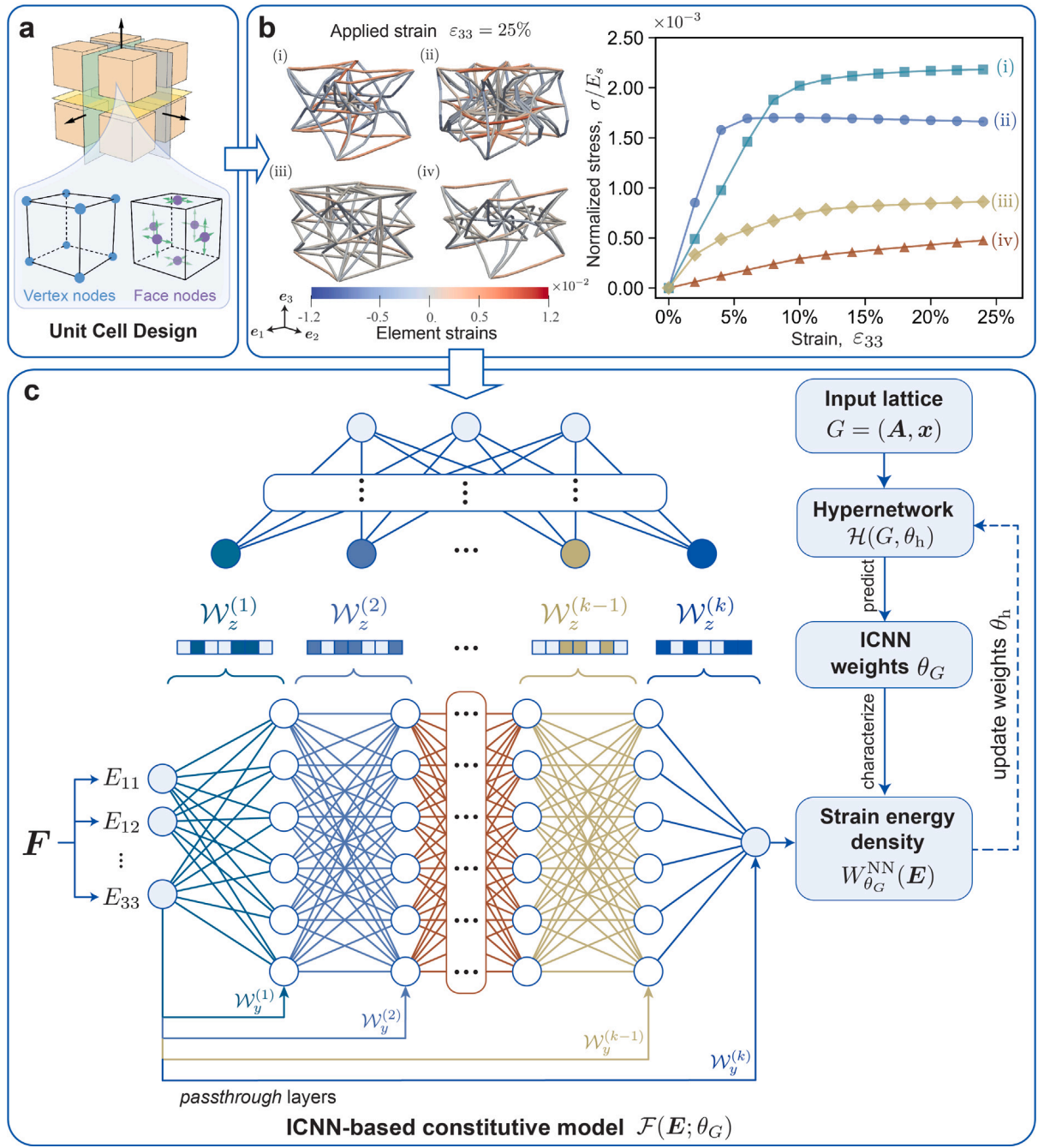

**Fig. 1.** Overview of the truss design space with nonlinear mechanical responses. **(a)** Cube decomposition generates the unit cell with possible node placements and the nodes' degrees of freedom defined on the octant. **(b)** Representative examples of deformed truss unit cells and their corresponding homogenized stress–strain responses under uniaxial compression along $e_3$ direction, showcasing the diverse design space of truss lattices with distinct deformation mechanisms. Reported stress values are normalized by the base material Young's modulus $E_s$. **(c)** Schematic of the HyperCAN framework for constitutive modeling of truss lattices. The hypernetwork $\mathcal{H}$ predicts the weight parameters $\theta_G$ that characterize an ICNN-based constitutive model for a given input graph representation of the truss lattice $G = (A, x)$. The constitutive model $\mathcal{F}(E; \theta_G)$ first calculates the Green–Lagrange strain measure $E$ for given deformation gradient $F$, and then predicts a convex strain energy density $W^{NN}_{\theta_G}$ with respect to the vectorized input $E$ through fully-connected layers and passthrough layers.

The schematic representation of the proposed HyperCAN framework is summarized in Fig. 1c. The target network, implemented as the ICNN-based constitutive model as described in Section 2.2, learns the strain energy density $W^{NN}_\theta$ while encoding necessary constitutive principles and preserving physical consistency. Instead of learning a static set of weights, the target network's parameters are generated on-the-fly by the hypernetwork conditioned on specific inputs, which allows the model to navigate the complex parameter spaces associated with various truss designs. That is, given an input truss lattice, represented by the graph $G = (A, x)$, the hypernetwork $\mathcal{H}$ predicts a customized constitutive model, denoted as $W_{\theta_G}^{\text{NN}}$, for the given truss as

$$\theta_G = \mathcal{H}(G; \theta_h) \quad \Rightarrow \quad W_{\theta_G}^{\text{NN}} = \mathcal{F}(E; \theta_G) = \mathcal{F}(E; \mathcal{H}(G; \theta_h)). \quad (14)$$

**Table 1**
**Overview of datasets for training and comprehensive evaluation of the HyperCAN model.** $\mathcal{D}^{\text{train}}$ is used for model training. $\mathcal{D}_L^{\text{test}}$ tests the model's capability to handle unseen loading scenarios on known truss structures, while $\mathcal{D}_G^{\text{test}}$ assesses generalization to entirely unseen truss structures under known loading conditions. $\mathcal{D}_{G,L}^{\text{test}}$ features both unseen truss structures and loading scenarios, aiming to test the model's adaptability to completely unknown conditions. $N$ is the number of unique truss structures, $K$ represents the number of unique loading paths, and $T$ is the number of loading steps per each. Deformation paths include UC (uniaxial compression), BC (biaxial compression), SS (simple shear), and TC (triaxial compression).

| Dataset | Truss structures | Loading scenarios | Deformation paths | Sample size | Dataset size |
| --- | --- | --- | --- | --- | --- |
| $\mathcal{D}^{\text{train}}$ | seen | seen | UC, BC, SS | N = 6000, K = 14, T = 100 | 6,134,800 |
| $\mathcal{D}_L^{\text{test}}$ | seen | unseen | UC, BC, SS, TC | N = 6000, K = 7, T = 100 | 3,163,800 |
| $\mathcal{D}_G^{\text{test}}$ | unseen | seen | UC, BC, SS | N = 3000, K = 14, T = 100 | 3,035,700 |
| $\mathcal{D}_{G,L}^{\text{test}}$ | unseen | unseen | UC, BC, TC | N = 200, K = 500, T = 100 | 4,344,800 |

The second Piola–Kirchhoff stress $S = \partial W/\partial E$ can be instantly obtained by differentiating the output of the predicted (target) constitutive NN, i.e., the strain energy density $W = W_{\theta_G}^{\text{NN}}$. By leveraging automatic differentiation [88] and backpropagation [89], the constitutive NN ensures smooth and analytically exact first and second derivatives, which is particularly beneficial for subsequent integration into FE simulations to improve numerical stability. The hypernetwork is designed as a feed-forward, multi-task learning architecture (see Appendix A for details). The MLP processes the graph representation of truss lattices as input, applies sequential transformations through several sharing hidden layers, and then predicts the weights for the fully connected layers and the passthrough layers of the (target) constitutive NN, respectively, via each task-specific layer. Here we consider a static target network architecture for our constitutive model, while the framework can be easily extended to other scenarios where the target network can have dynamic architectures to maintain more flexibility [73].

Given a representative dataset with $N$ tuples of truss graph description $G$, Green–Lagrange strain tensors $E$, second Piola–Kirchhoff stress tensors $S$, and strain energy densities $W$ (all measured along $K$ loading paths, each discretized into $T$ load steps) as

$$\mathcal{D} = \left\{ \left( G^{(n)}, E_t^{(n,k)}, S_t^{(n,k)}, W_t^{(n,k)} \right) : n = 1, \ldots, N; k = 1, \ldots, K; t = 1, \ldots, T \right\}, \quad (15)$$

the hypernetwork $\mathcal{H}$ is trained to minimize the loss function as

$$\theta_h \leftarrow \arg\min_{\theta_h} \lambda_S \underbrace{\frac{1}{|\mathcal{D}|} \sum_{n=1}^{N} \sum_{k=1}^{K} \sum_{t=1}^{T} \|S_t^{(n,k)} - \hat{S}_t^{(n,k)}\|^2}_{\text{stress prediction loss}}$$
$$+ \lambda_W \underbrace{\frac{1}{|\mathcal{D}|} \sum_{n=1}^{N} \sum_{k=1}^{K} \sum_{t=1}^{T} \|W_t^{(n,k)} - \hat{W}_t^{(n,k)}\|^2}_{\text{energy prediction loss}}, \quad (16)$$

with $\quad \hat{W}_t^{(n,k)} = W_{\theta_G}^{\text{NN}}\left(E_t^{(n,k)}\right) \quad$ and $\quad \hat{S}_t^{(n,k)} = \partial \hat{W}_t^{(n,k)} / \partial E_t^{(n,k)},$

where $n, k,$ and $t$ denote the indices for different truss structures, deformation paths, and loading steps, respectively. $\lambda_S > 0$ and $\lambda_W > 0$ are the weights for the stress prediction loss and the energy prediction loss, respectively. We choose to include the energy prediction loss with a smaller weight ($\lambda_S = 1, \lambda_W = 0.2$) as a regularization term, guiding the model towards predictions that are not only accurate in terms of stress predictions but also physically plausible in terms of energy landscape consistent with the stress responses, thereby enhancing the model's ability to extrapolate to unseen deformation states. Details regarding the selection of the loss weights are presented in Appendix B. By learning the mapping from truss configurations to constitutive model weights, the hypernetwork encapsulates distinct material behaviors within a unified framework, which enables efficient training and inference for a wide range of truss metamaterials and, furthermore, enhances the model's generalization and extrapolation ability to unseen truss structures without the need for retraining from scratch for each new configuration.

We train our HyperCAN framework with the training dataset $\mathcal{D}^{\text{train}}$, containing $K = 14$ different loading paths for 6,000 unique truss structures, with $T = 100$ loading steps for each load path. To systematically evaluate the adaptability and robustness of our HyperCAN model, we consider several test datasets, as outlined in Table 1. Specifically, $\mathcal{D}_L^{\text{test}}$ tests the model's ability to handle unseen loading scenarios for known truss structures, whereas $\mathcal{D}_G^{\text{test}}$ evaluates the model's generalization to unseen truss structures under familiar loading conditions. Note that, although $\mathcal{D}_L^{\text{test}}$ contains deformation paths {UC, BC, SS} that are also present in training, specific loading parameters within these paths are not seen during training, which ensures that the model is tested on truly unseen scenarios. Furthermore, we randomly sample 500 deformation paths including UC, BC, and TC with uniformly sampled loading parameters $\lambda_1, \lambda_2, \lambda_3 \in [-0.3, 0]$ (extending beyond those sampled in the training dataset). The constructed dataset $\mathcal{D}_{G,L}^{\text{test}}$, composed exclusively of *unseen* loading paths applied to *unseen* truss designs, aims to provide a more stringent test of the constitutive model's extrapolation capabilities to completely unknown conditions. All details pertaining to the NN architectures, training protocols, and hyperparameters are presented in Appendix A.

## 3. Results

### 3.1. Performance of the constitutive model

We first examine our model's predictive performance on stress–strain responses for truss structures along unseen deformation paths from $\mathcal{D}_L^{\text{test}}$. Reported stress values are normalized by the base material's Young's modulus $E_s$. Fig. 4a presents the accuracy of the model for the dataset $\mathcal{D}_L^{\text{test}}$, demonstrated by the overall goodness-of-fit $R^2 \geq 99.4\%$ across all normal stress components (see Appendix B for detailed results). For this and subsequent $R^2$-plots, stress values for each type of deformation path are normalized by their corresponding standard deviation for better visualization, given the significant variation in stress magnitudes across different deformations, such as simple shear and biaxial compression, as is commonly the case for anisotropic materials. Fig. 2 compares the predicted vs. true stress responses for four representative examples from $\mathcal{D}_L^{\text{test}}$ subject to biaxial compression and triaxial compression deformations, exhibiting distinct nonlinear behaviors (see also Supplementary Movies 1–3). The constitutive model $W_{\theta}^{\text{NN}}$ accurately predicts the mechanical responses for these unseen loading paths, achieving an averaged normalized root mean square error (NRMSE) of 1.35% (see Appendix B for detailed results). Altogether, this confirms that, by embedding fundamental principles of mechanics and material behavior directly into the model architecture, our constitutive model can not only capture the intricate mechanical behavior of truss metamaterials under various nonlinear loading conditions but also generalize effectively to unseen deformation paths.

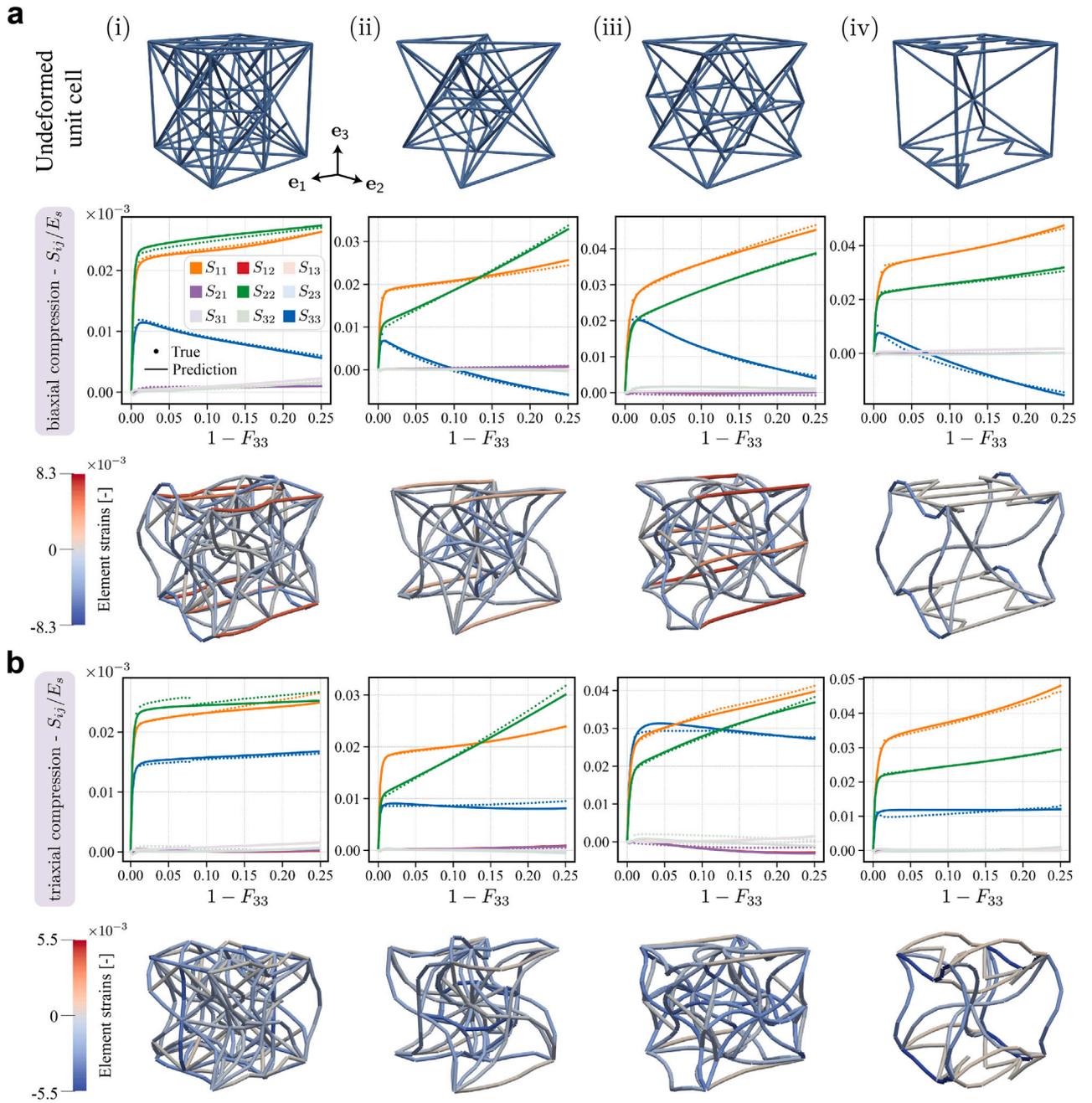

**Fig. 2.** Evaluation of constitutive model predictions on stress responses for unseen loading scenarios. Stress–strain plots showing the predicted (solid lines) vs. true (solid dots) responses for representative truss structures taken from $\mathcal{D}_L^{\text{test}}$ along **(a)** biaxial compression and **(b)** triaxial compression deformation paths sampled from outside of the training dataset. The resulting deformed unit cells are also shown alongside. Reported stress values are normalized by the base material Young's modulus $E_s$.

Next, we assess the performance of the HyperCAN on $\mathcal{D}_G^{\text{test}}$, which evaluates the model's generalization to unit cells not encountered during training. For each structure, the HyperCAN instantly predicts a constitutive model with its weights tailored to the specific truss topological and geometrical features, which allows the ML model to learn and share information from existing domains and adapt effectively to unknown domains. Fig. 4b shows the accuracy of the predicted stress responses for the dataset $\mathcal{D}_G^{\text{test}}$, with $R^2 \geq 99.1\%$ across all normal stress components (see Appendix B for detailed results). Fig. 3 illustrates four representative unit cells sampled from the holdout set $\mathcal{D}_G^{\text{test}}$, comparing their true vs. predicted stress responses subjected to simple shear and biaxial compression (see also Supplementary Movies 4–5). The model achieves an averaged NRMSE of 2.18% on unseen structures, demonstrating robust generalization to novel designs. We observe that the unit cells shown in Fig. 2b(i) and Fig. 3b(ii) exhibit notable similarities by visual inspection, while their corresponding mechanical responses differ significantly. This highlights the challenges in developing unified constitutive models for a broad range of structural designs, which can accommodate the high sensitivity of the mechanical behavior to structural features. Overall, this underscores HyperCAN's efficacy in learning the underlying intricate dependencies between metamaterial design and constitutive behavior for a broad truss design space, going beyond those designs previously encountered during training.

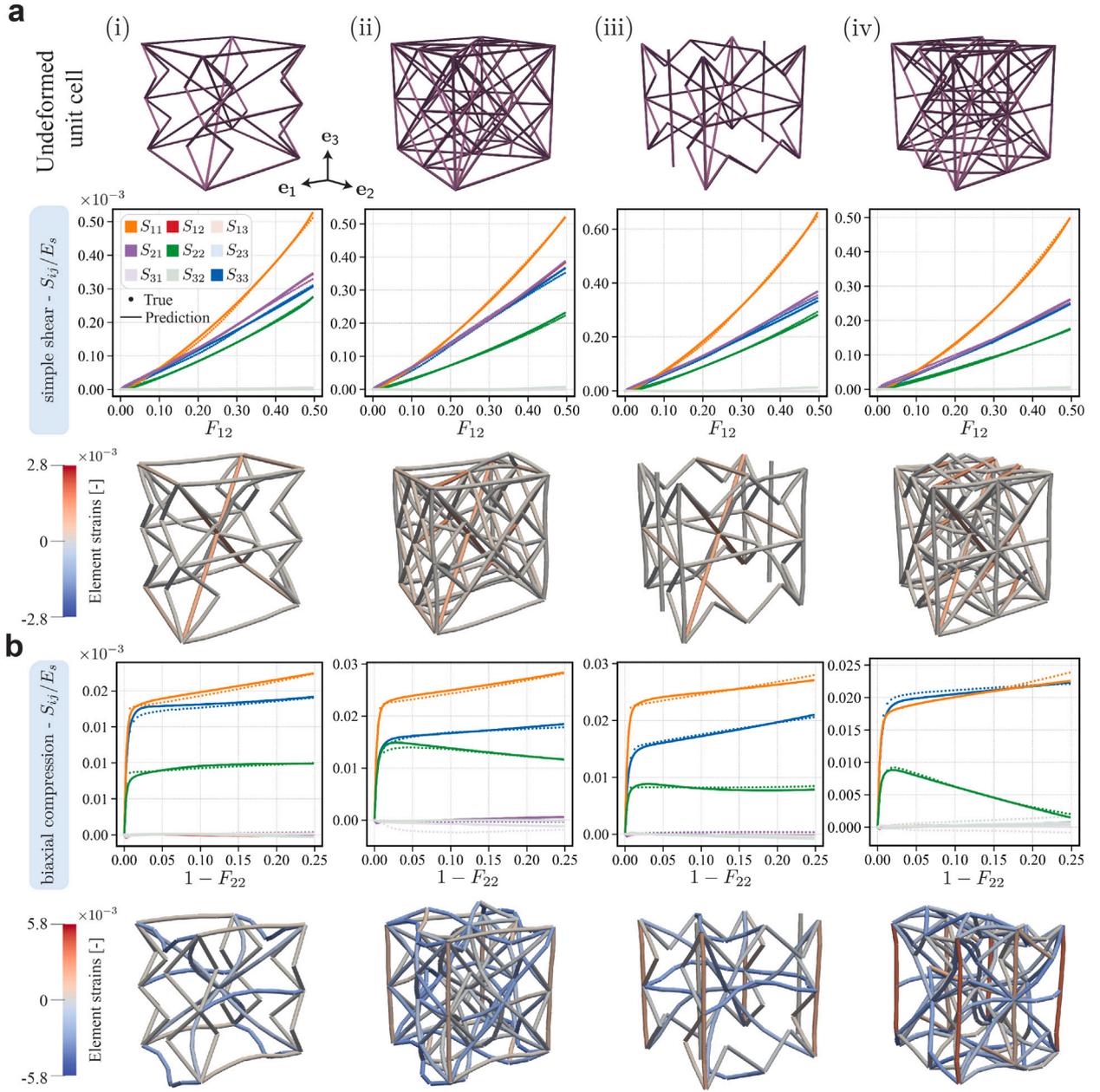

**Fig. 3.** Evaluation of constitutive model predictions on stress responses for unseen truss structures. Stress–strain plots showing the predicted (solid lines) vs. true (solid dots) responses for representative truss structures taken from $\mathcal{D}_G^{\text{test}}$ (i.e., not included during training) along **(a)** simple shear and **(b)** biaxial compression deformation paths. The resulting deformed unit cells are also shown alongside. Reported stress values are normalized by the base material Young's modulus $E_s$.

### 3.2. Generalization beyond the training domain

To further validate the model's versatility, we evaluate the model's predictive power for unseen designs with highly anisotropic properties, especially when going to large deformations beyond the training range. The dataset $\mathcal{D}_{G,L}^{\text{test}}$, as outlined in Table 1, presents the most challenging test case, featuring both unseen truss structures and unseen loading scenarios with randomly sampled loading parameters, which serves as a rigorous assessment of the model's effectiveness in entirely unknown domains. To visualize the model's generalization capabilities, Fig. 5a compares different loading scenarios in the strain invariants space for both the training (seen) and evaluation (unseen) deformation paths

across all loading steps. The principal strain invariants are computed as

$$\bar{I}_1 = J^{-2/3} I_1, \quad \bar{I}_2 = J^{-4/3} I_2, \quad J = \det \mathbf{F} = I_3^{1/2}, \quad (17)$$

with $I_1 = \operatorname{tr} \mathbf{C}, \quad I_2 = \frac{1}{2}\left[(\operatorname{tr} \mathbf{C})^2 - \operatorname{tr}(\mathbf{C}^2)\right], \quad I_3 = \det \mathbf{C},$

where $\mathbf{C} = \mathbf{F}^T \mathbf{F} = \mathbf{I} + 2\mathbf{E}$ is the right Cauchy–Green tensor. Additionally, we highlight two representative test deformation paths in Fig. 5 and evaluate the model's predictions on unseen truss structures (taken from $\mathcal{D}_G^{\text{test}}$, consistent with Fig. 3) subject to these test paths. Fig. 5 highlights that the unseen deformation states used in $\mathcal{D}_L^{\text{test}}$ and $\mathcal{D}_{G,L}^{\text{teset}}$ considerably surpass the boundaries of those encountered during the model's training phase, showcasing the extensive range of the model's applicability. As shown in Fig. 5b, the model demonstrates remarkable accuracy in predicting responses at unseen deformation states that extend significantly beyond the training space (see also

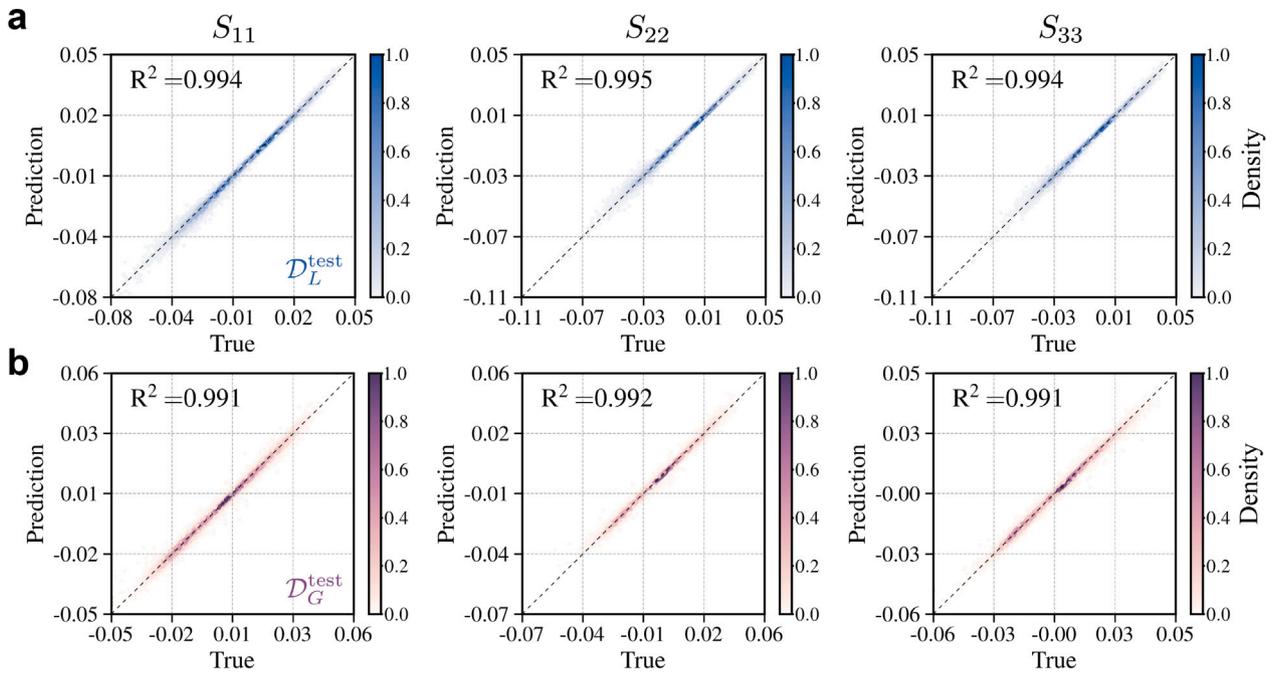

**Fig. 4.** Evaluation of constitutive model prediction accuracy. Distribution of predicted vs. ground truth stress responses (obtained by FE homogenization) for **(a)** unseen loading scenarios $\mathcal{D}_L^{\text{test}}$ and **(b)** unseen truss structures $\mathcal{D}_G^{\text{test}}$ for the normal components $S_{11}$, $S_{22}$, and $S_{33}$ of the second Piola–Kirchhoff stress tensor (see Appendix B for detailed results). The color of each marker indicates the normalized density of the corresponding data point over the dataset distribution. Dashed lines denote ideal lines with zero intercept and unit slope; the goodness-of-fit $R^2$ is shown as a measure of accuracy.

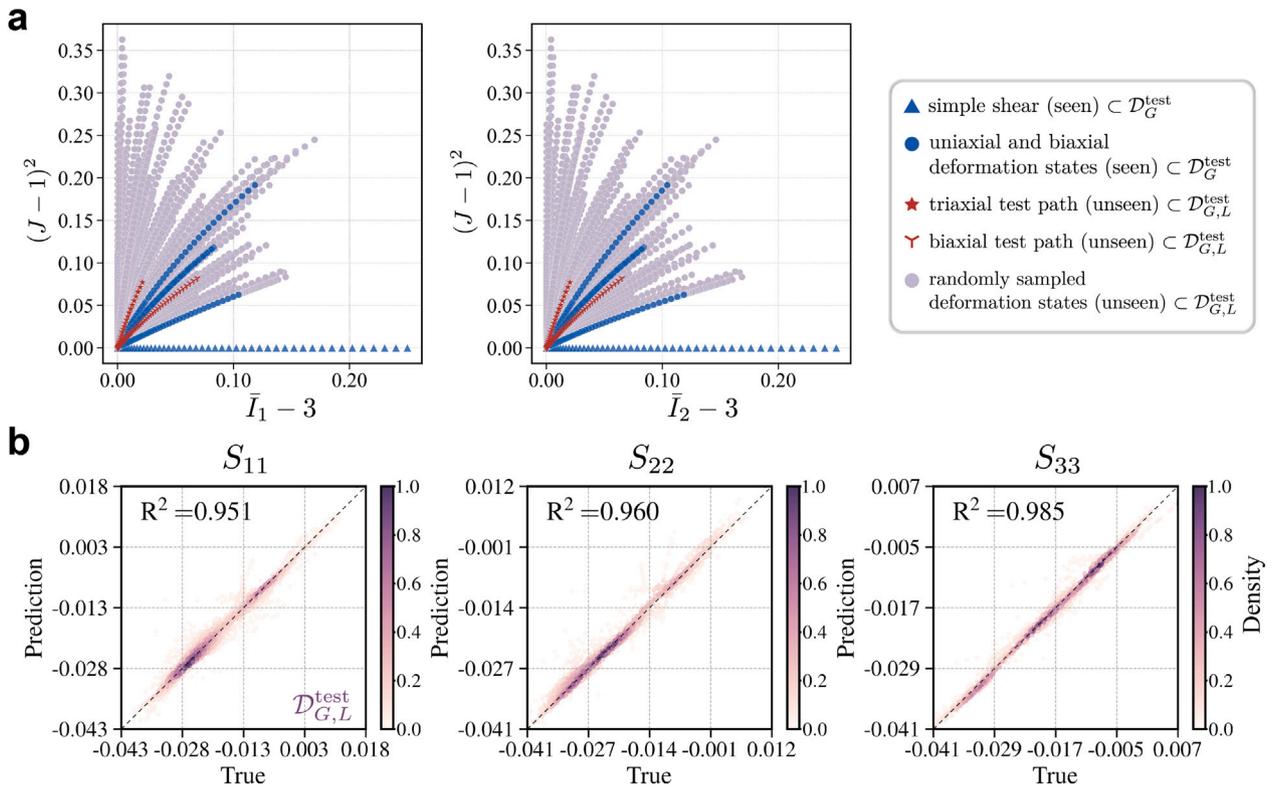

**Fig. 5.** Evaluation of constitutive model predictions beyond the training domain. **(a)** Principal strain invariant components $\bar{I}_1 - 3$, $\bar{I}_2 - 3$, and $(J-1)^2$ at each deformations state used in training (seen) and evaluation (unseen) across all loading steps. Also shown are the principal strain invariants of the biaxial test and triaxial test deformation path used in Fig. 6b across all loading steps. **(b)** Distribution of predicted vs. ground truth stress responses for unseen truss structures along unseen deformation paths $\mathcal{D}_{G,L}^{\text{test}}$ for the normal components $S_{11}$, $S_{22}$, and $S_{33}$ of the second Piola–Kirchhoff stress tensor (see Appendix B for detailed results). The color of each marker indicates the normalized density of the corresponding data point over the dataset distribution. Dashed lines denote ideal lines with zero intercept and unit slope; the goodness-of-fit $R^2$ is shown as a measure of accuracy.

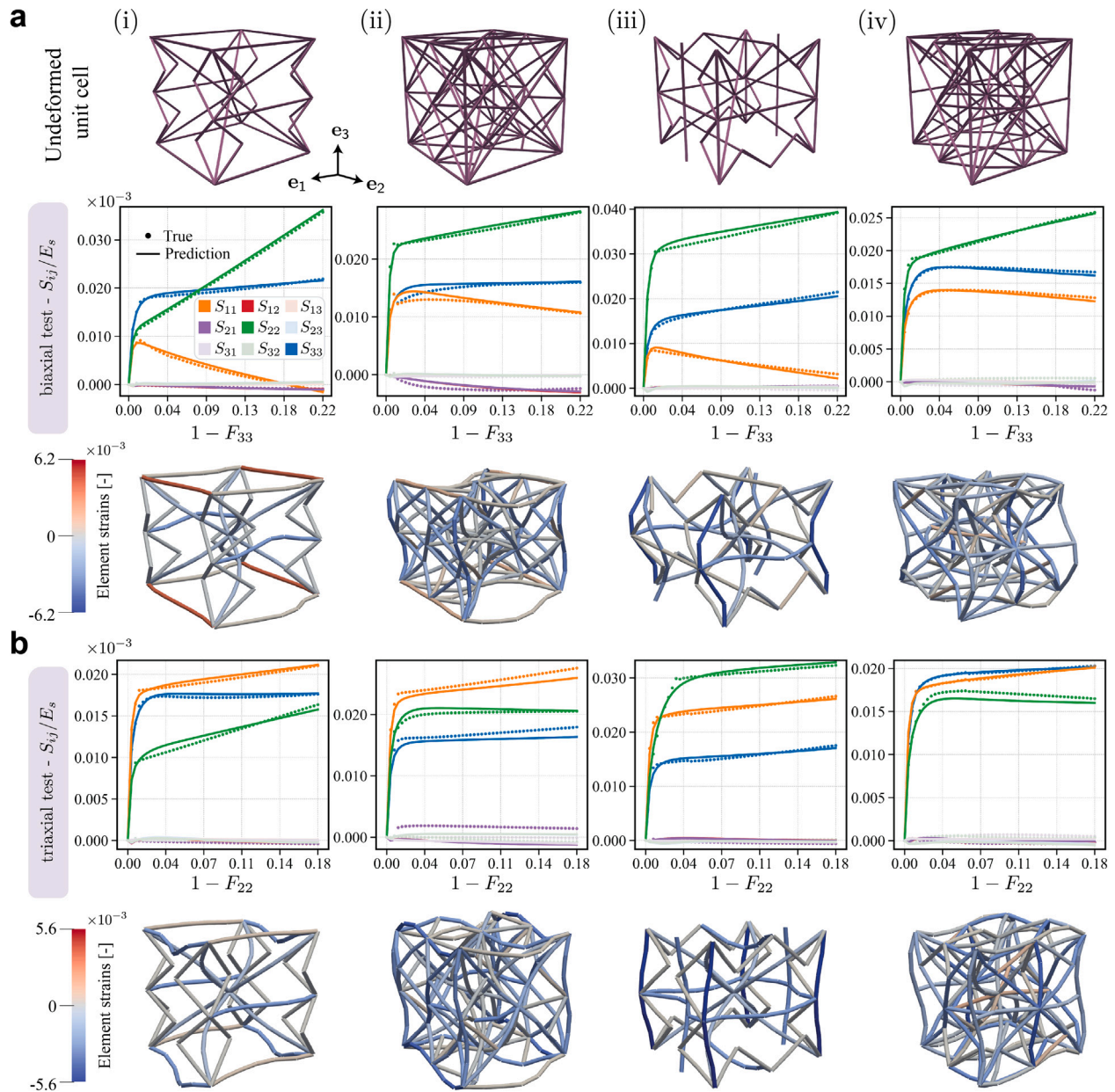

**Fig. 6.** Evaluation of constitutive model predictions on stress responses for unseen unit cells along unseen deformation paths. Stress–strain plots showing the predicted (solid dots) vs. true (solid lines) responses for representative truss structures taken from $\mathcal{D}_{G,L}^{\text{test}}$ (i.e., not included during training) along **(a)** simple shear and **(b)** biaxial compression deformation paths. The resulting deformed unit cells are also shown alongside. Reported stress values are normalized by the base material Young's modulus $E_s$.

Supplementary Movies 6–7), achieving an averaged NRMSE of 5.23% and $R^2 \geq 95.2\%$ across all unseen test cases (see Appendix B for detailed results). Furthermore, Fig. 6 shows that the model's prediction agrees well with the ground truth for stress responses of unseen truss unit cells along the unseen loading paths, demonstrating the model's robust generalization capabilities in handling completely unseen scenarios.

### 3.3. Multiscale simulations with hypercan

HyperCAN's ability to generalize to unseen load conditions makes it a promising tool for efficient multiscale simulations of architected materials. One of the main challenges in modeling their highly nonlinear and anisotropic mechanical response is the identification of an effective constitutive model—especially when considering geometric nonlinearity and instability at finite strains. If such an effective material model is available, it can be used efficiently in FE simulations that replace the multitude of discrete structural elements by an effective continuum. The challenge in this setting is the appearance of locally arbitrary deformation histories in the heterogeneously deforming continuum, for which the effective constitutive model must be suitable. To demonstrate the applicability of our framework, we employ HyperCAN within a 3D multiscale FE simulation framework to characterize the global mechanical responses, where the constitutive NN serves as a surrogate model to provide the local constitutive law of the macroscale body, offering a computational shortcut to expensive, fully resolved structural simulations.

We consider a macroscale continuous body $\Omega \subset \mathbb{R}^3$, whose underlying microstructure is the periodic tessellation of an RVE. The RVE, in turn, is composed of struts modeled as linear elastic corotational (Euler–Bernoulli) beam elements. In a classical $FE^2$ setting, for every macroscale point $x \in \Omega$ the effective strain energy density $W(E)$ is computed on the fly from a homogenization simulation on the RVE-level [90]. As an efficient alternative, we here use $W_{\theta_G}^{\text{NN}}(E)$ as the effective strain energy density on the macroscale.

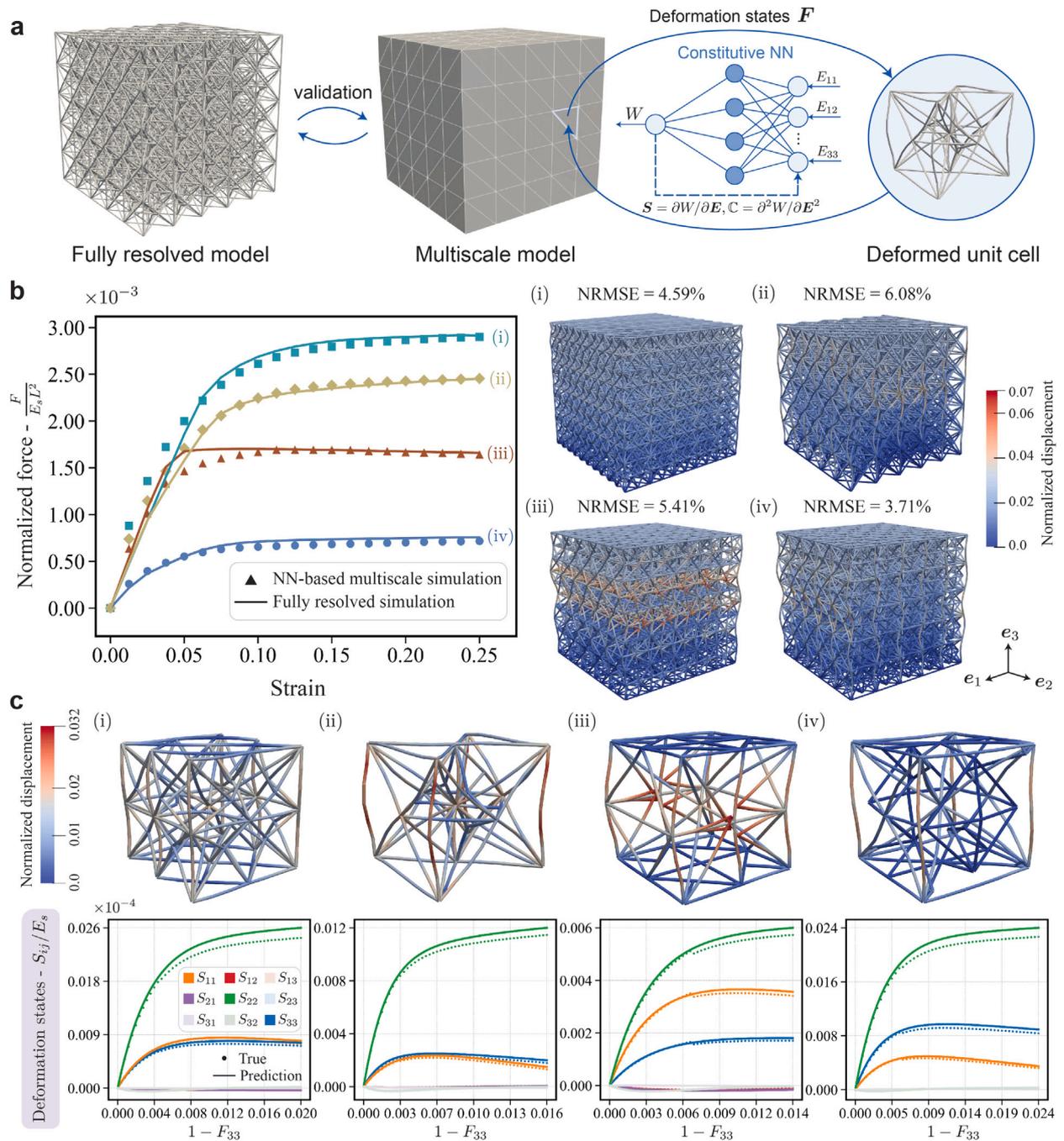

**Fig. 7.** Data-driven nonlinear multiscale simulation framework for truss metamaterials. **(a)** The discrete structure is simulated as a continuous body, which is discretized by tetrahedral finite elements. At every integration point of the FE mesh, the effective constitutive behavior of a unit cell is predicted by the NN model, taking the (local) macroscopic deformation field $F$ as input. The predicted RVE-averaged stresses $S$ and tangent modulus $\mathbb{C}$ are calculated by differentiating the energy and returned to the macroscale. **(b)** Global mechanical responses (normalized by the base material Young's modulus $E_s$ and the cube side length $L$) calculated by the NN-based continuum model are compared with those obtained from the fully resolved discrete beam model for four representative structures made of truss unit cells taken from $\mathcal{D}^{\text{test}}_{G,L}$ to validate our approach. Each of the four examples shows the corresponding deformed structure obtained by the fully resolved model and its displacement magnitude (normalized by the cube side length $L$) under an applied compressive displacement of 0.025 m. Also shown are the NRMSE of the compressive forces computed by the fully resolved model and the continuum model. **(c)** Deformed truss unit cell for each structure and its displacement magnitude (normalized by the unit cell dimension) are visualized at a compressive displacement of 0.025 m. Stress–strain plots compare the predicted (solid lines) vs. true (solid dots) homogenized responses of the four structures at the mesh integration point highlighted in (a) along their corresponding deformation paths.

Fig. 7a shows a schematic representation of the NN-based multiscale simulation framework for beam-based metamaterials. For a given architecture $G = (A, x)$, the hypernetwork predicts a constitutive NN that serves as a surrogate model for the effective strain energy density. The effective second Piola–Kirchhoff stress tensor $S$ and the tangent stiffness tensor $\mathbb{C}$ are computed efficiently from the NN model by automatic differentiation via Eq. (8), which are then returned to the macroscale. The macroscale boundary value problem is solved by the FE method, which discretizes the macroscale body into tetrahedral elements. The resulting nonlinear system of equilibrium equations is solved by Newton–Raphson iteration.

As an example, we consider a cubic sample of size $0.1 \times 0.1 \times 0.1 \text{m}^3$ composed of $5 \times 5 \times 5$ unit cells under uniaxial compression. Within each unit cell, beams are assumed to have constant circular cross-sections characterized by an isotropic linear elastic base material with Young's modulus $E_s = 1$ and Poisson's ratio $\nu = 0.3$. Compressive displacements up to $u = -0.025$m are applied incrementally across the top surface in the $e_3$–direction, while the bottom surface is fixed and the other surfaces remain unconstrained. The deformation within the sample is not homogeneous in general and can show buckling instability. We employ the aforementioned NN-based multiscale simulation approach to study the global mechanical response of large truss structures made of different unit cell designs. Results are compared with a 3D fully resolved beam model to illustrate the validity of our NN-based continuum model.

Fig. 7b showcases four representative examples of deformed truss structures made of $5 \times 5 \times 5$ truss unit cells, and their corresponding overall force–displacement responses obtained from both the discrete fully resolved and NN-based multiscale simulations. The compression force is normalized by the base material's Young's modulus $E_s$ and the side length of the cube $L$. Fig. 7c further illustrates the deformed truss unit cell for each structure, and additionally, the constitutive NN predictions for their homogenized responses along the corresponding loading path for the element integration point in the macroscale model shown in Fig. 7a (see also Supplementary Movie 8). The NN-based homogenized model reveals good agreement with the fully resolved model (with an overall NRMSE $\leq 6.08\%$) across the nonlinear deformation regime, which confirms that the NN model effectively captures the various truss responses at large deformation. In Fig. 7b(i) and (ii), we observe steady increases in force which eventually plateaus at large displacements, while the samples in Fig. 7b(iv) and especially (iii) demonstrate distinct softening, which indicates buckling instabilities within the structure (of individual unit cells or rows of cells). Notably, the NN constitutive model accurately represents such phenomena in the global mechanical responses of large truss structures, as illustrated in Fig. 7b(iii) and (iv). This validation underscores the robustness and the generalization capability of the NN constitutive model in predicting the complex mechanical responses of beam lattices including buckling instabilities.

The close agreement of the NN-based homogenized models with the fully resolved simulation across various truss structures demonstrates the potential of HyperCAN to serve as an efficient surrogate in large-scale simulations, offering a cost-effective alternative to fully resolved structural calculations with orders of magnitude improvement in computational efficiency (see Appendix C for detailed comparisons of computational performance). Furthermore, the NN constitutive model inherently ensures smooth, continuous predictions and derivatives without spurious oscillations, e.g., as shown in Fig. 7c(iii). This smoothness, while potentially smoothing out abrupt changes in mechanical responses due to, e.g., localized buckling effects, offers significant advantages when employed in FE simulations under large deformations by enhancing computational efficiency and numerical stability. Evaluation of the once-trained NN-based model is considerably more efficient than on-the-fly homogenization [90] (see Table C.4), while also generalizing to metamaterial architectures, which offers opportunities for multiscale optimization [42,91].

## 4. Conclusion

The mechanical modeling of beam-based architected materials, featuring a vast topological and geometrical design space, presents challenges due to geometric nonlinearity and instability under large deformations. Here, we introduce HyperCAN, a unified framework for constructing parameter-efficient, mechanics-informed, and highly adaptable constitutive models for a broad spectrum of beam lattices exhibiting distinct nonlinear mechanical behaviors. The proposed NN constitutive models demonstrate remarkable robustness and generalization to unseen unit cell designs and unseen loading scenarios outside the training domain, further enabling efficient multiscale simulations that replace discrete structures by effective continua.

The HyperCAN framework tackles the challenges of learning constitutive NNs for different designs by employing hypernetworks, which dynamically adjust the constitutive relations (via the weights and biases of a constitutive NN) to a given design. This decomposition bypasses the need to store all structure–property relations within the constitutive model, thereby facilitating efficient learning and transferring to novel structures. We construct the constitutive models using ICNNs, which inherently fulfill thermodynamic consistency, local material stability, a stress-free reference configuration, and objectivity. Integrating fundamental physical principles and mechanics considerations into the constitutive NN significantly enhances the model's performance, especially the generalization to unseen strain states. Furthermore, the ICNN avoids non-smooth, oscillatory predictions of the stress response, ensuring smooth first and second derivatives of the effective strain energy density. This admits the incorporation of the constitutive NN into nonlinear multiscale simulations.

The presented constitutive modeling framework closes the gap in constructing efficient, physically sensible mappings between structural configurations to finite-strain responses for a wide class of beam-based architected materials. The graph-based representation with great versatility provides a unified parameterization for capturing complex interactions among topological features of truss unit cells [19], facilitating the exploration of a wide range of potential microstructure designs in multiscale design and optimization of architected materials. It also admits straightforward extension to more extensive design space of trusses, accommodating variations in, e.g., relative density or manufacturing imperfections [92], as well as to other classes of metamaterials, e.g., triply periodic minimal surface (TPMS) lattices [93,94] or spinodal morphologies [5,6,95,96] by modifying the design space parameterization. Moreover, the proposed framework can be adapted to address more complex material behavior, e.g., path- or history-dependent base material models [97], ranging from viscoelasticity and plasticity, by modifying the core architecture of constitutive NNs. For example, hypernetworks can be leveraged to generate parametric physics-informed neural networks (PINNs) that approximate the solutions to the underlying partial differential equation for a given parameterization [98]. This highlights the robustness and versatility of our HyperCAN framework, opening up new avenues for the multiscale design and analysis of a wide range of architected materials.

## CRediT authorship contribution statement

**Li Zheng:** Writing – original draft, Visualization, Validation, Software, Methodology, Data curation. **Dennis M. Kochmann:** Writing – review & editing, Supervision, Methodology, Conceptualization. **Siddhant Kumar:** Writing – review & editing, Supervision, Methodology, Conceptualization.

## Code availability

The code used to train and evaluate the model is available at https://github.com/li-zhengz/HyperCAN and on Zenodo [99].

## Declaration of competing interest

The authors declare that they have no known competing financial interests or personal relationships that could have appeared to influence the work reported in this paper.

## Acknowledgments

L.Z. was supported by ETH Zurich through the SynMatLab project.

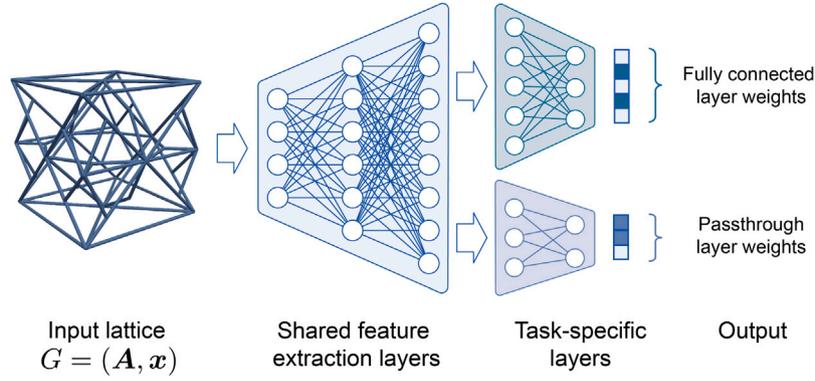

**Fig. A.8.** Schematic illustration of the multi-tasking hypernetwork architecture. The multi-tasking hypernetwork first processes the input truss graph representation through several shared feature extraction layers, and then feeds the intermediate outputs into separate task-specific branches corresponding to the prediction of fully connected layer weights and the passthrough layer weights of the ICNN constitutive model.

## Appendix A. Model architecture and training protocols

In this work, we employ a multi-layer perceptron (MLP) to construct the hypernetwork $\mathcal{H}$, which predicts the weights of the ICNN constitutive model based on given input truss graph representations $G = (\boldsymbol{A}, \boldsymbol{x})$. We note that since the adjacency matrix $\boldsymbol{A}$ is symmetric, only the upper triangular part is considered by the model. Additionally, our focus is restricted to the node positions $\boldsymbol{x}$ of 6 face nodes $\{f_1, f_2, f_3, f_4, f_5, f_6\}$, as positions of other vertex nodes remain constant. Given the distinct functionalities and constraints on the weights of the fully connected and passthrough layers in the ICNN architecture, we adopt a multi-task architecture for the hypernetwork, where the intermediate outputs of the shared layers are fed into separate blocks of task-specific layers to predict the weights of different layers. A schematic representation of this multi-tasking hypernetwork architecture is depicted in Fig. A.8. Further details of the relevant hyperparameters of the HyperCAN framework, including the hypernetwork and the ICNN-based constitutive model, are provided in Table A.2. In Algorithm 1, we summarize a pseudocode for the training of the HyperCAN framework.

In our framework, the hypernetwork and ICNN are trained jointly in an end-to-end manner without any restrictions on their interaction. Specifically, the hypernetwork processes the graph-based representation of the truss configuration and generates the weights for the ICNN, and as such, there is no independent learning process for the ICNN itself. The ICNN then uses these weights to predict the strain energy density and compute the stress responses by differentiating the energy with respect to the input deformation measures. The ICNN's weights are entirely determined by the hypernetwork based on the input truss configurations, and consequently, the hypernetwork is the only component that is trained and optimized. During training, the gradients from the stress and energy prediction losses propagate through the ICNN back to the hypernetwork, which are then used to update the hypernetwork's parameters. This allows the hypernetwork to directly adjust the ICNN weights in response to the error between the predicted mechanical responses and the ground truth data.

The joint training process enables the model to effectively capture the complex mapping from truss configurations to mechanical responses in a unified optimization framework – eliminating the need for separate training stages for each component. Furthermore, this approach facilitates better generalization to unseen truss structures, as the hypernetwork learns to dynamically adjust the ICNN's behavior for a wide range of truss configurations and deformation scenarios without requiring retraining for new domains.

## Appendix B. Model performance

Fig. B.9 shows the evolution of the loss and the normalized root mean square error (NRMSE, computed as in Eq. (B.1)) against the

---

**Algorithm 1** HyperCAN training

**Input:** Graph representation of truss lattice $G = (\boldsymbol{A}, \boldsymbol{x})$
**Input:** Deformation gradient $\boldsymbol{F}$, strain energy density $W$, and Second Piola-Kirchhoff stress $\boldsymbol{S}$

1: Randomly initialize hypernetwork parameters: $\theta_h$.
2: Initialize Adam optimizer with learning rate $\eta$.
3: **for** $e = 1, \ldots, n_e$ **do** ▷ Iterate over training epochs
4:   Loss: $\mathcal{L} \leftarrow 0$ ▷ Initialize loss
5:   $\theta_G : \{(\mathcal{W}_z^{(i)}, \mathcal{W}_y^{(i)}) : i = 1, \ldots, k\} \leftarrow \mathcal{H}(G; \theta_h)$ ▷ Compute layer weights of the ICNN
6:   $\boldsymbol{E} \leftarrow \frac{1}{2}(\boldsymbol{F}^T \boldsymbol{F} - \boldsymbol{I})$ ▷ Compute Green-Lagrangian strain tensor
7:   $\boldsymbol{y} \leftarrow [E_{11}, E_{12}, E_{13}, E_{21}, E_{22}, E_{23}, E_{31}, E_{32}, E_{33}]^T$ ▷ Vectorized input
8:   $\boldsymbol{z}^{(1)} \leftarrow \phi\left(\mathcal{W}_y^{(1)} \boldsymbol{y} + \boldsymbol{b}^{(1)}\right)$ ▷ Input layer of the ICNN
9:   **for** $i = 2, \ldots, k-1$ **do** ▷ Iterate over ICNN hidden layers
10:    $\boldsymbol{z}^{(i)} \leftarrow \phi\left(\sigma(\mathcal{W}_z^{(i)})\boldsymbol{z}^{(i-1)} + \mathcal{W}_y^{(i)} \boldsymbol{y} + \boldsymbol{b}^{(i)}\right)$
11:   **end for**
12:   $\hat{W}_{\theta_G} \leftarrow \sigma(\mathcal{W}_z^{(k)})\boldsymbol{z}^{(k-1)} + \mathcal{W}_y^{(k)} \boldsymbol{y} + \boldsymbol{b}^{(k)}$ ▷ Compute strain energy density
13:   $\hat{\boldsymbol{S}} \leftarrow \partial \hat{W}_{\theta_G} / \partial \boldsymbol{E}$ ▷ Compute Second Piola-Kirchhoff stress
14:   $\mathcal{L} \leftarrow \mathcal{L} + \lambda_S (\hat{\boldsymbol{S}} - \boldsymbol{S})^2 + \lambda_W (\hat{W}_{\theta_G} - W)^2$ ▷ Update loss
15:   Compute gradients $\nabla_{\theta_h} \mathcal{L}$ using automatic differentiation.
16:   Update parameters $\theta_h$ using the Adam optimizer:
17:    $\theta_h \leftarrow \text{Adam}(\theta_h, \nabla_{\theta_h} \mathcal{L}, \eta)$
18: **end for**
19: **Output:** Trained hypernetwork $\mathcal{H}(G; \theta_h)$; constitutive models $\hat{W}_{\theta_G} = \mathcal{F}(\boldsymbol{E}; \mathcal{H}(G; \theta_h))$

---

number of training iterations. We further show the error maps of NRMSE measures for each stress component for datasets $\mathcal{D}_L^{\text{test}}$, $\mathcal{D}_G^{\text{test}}$ and $\mathcal{D}_{G,L}^{\text{test}}$ in Fig. B.10, corresponding to Fig. 4a, Fig. 4b, and Fig. 5b in the main article, respectively.

$$\text{RMSE}(\boldsymbol{S}, \hat{\boldsymbol{S}}) = \sqrt{\frac{1}{n}\sum_{i=1}^{n} \|\boldsymbol{S}^{(i)} - \hat{\boldsymbol{S}}^{(i)}\|^2}, \quad \text{NRMSE}(\boldsymbol{S}, \hat{\boldsymbol{S}}) = \frac{\text{RMSE}(\boldsymbol{S}, \hat{\boldsymbol{S}})}{\max(\boldsymbol{S}) - \min(\boldsymbol{S})}$$

(B.1)

where $\boldsymbol{S} \in \mathbb{R}^9$ and $\hat{\boldsymbol{S}} \in \mathbb{R}^9$ denote the true and the predicted second Piola–Kirchhoff stresses, respectively, and $n$ denote the size of the dataset.

Besides, we have investigated the effects of different values of energy prediction loss weight $\lambda_W$ on the model's prediction performance. We choose to include the energy prediction loss in Eq. (16)

**Table A.2**
Dimensions and training hyperparameters of the HyperCAN ML framework.

| | Hypernetwork $\mathcal{H}$ | | | Constitutive model $\mathcal{F}_\omega$ |
|---|---|---|---|---|
| | Shared layers | Fully-connected layer weights predictor | Passthrough layer weights predictor | |
| Input dimensions | 117 | 256 | 256 | 9 |
| Hidden dimensions | 256,256,256,256 | 256,256,256,256 | 256,256,256,256 | 20,20,20 |
| Output dimensions | 256 | 1000 | 369 | 1 |
| Activation functions | | LeakyReLU [100] (with a negative slope of 0.01) | | – |
| Optimization algorithm | | Adam [101] | | – |
| Learning rate | | $5 \times 10^{-4}$ | | – |
| Batch size | | 64 | | – |
| Number of epochs | | 1000 | | – |

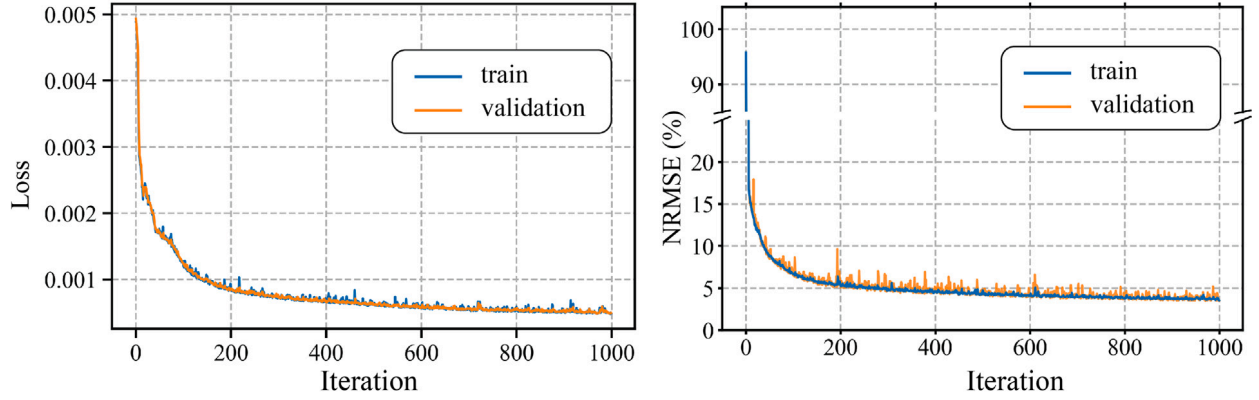

**Fig. B.9.** Error convergence evaluated across the entire training and validation dataset. Evolution of the **(a)** loss and **(b)** normalized root mean square error (NRMSE, expressed as a percentage) as a function of the number of training iterations.

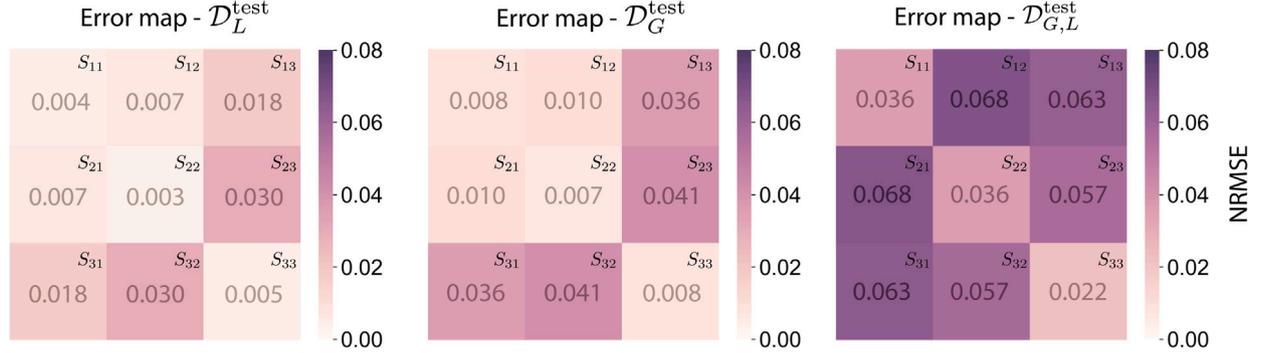

**Fig. B.10.** Error map of the normalized root mean square error (NRMSE) evaluated across different datasets. $\mathcal{D}_L^{\text{test}}$ evaluates the model's performance on unseen loading scenarios with known truss structures. $\mathcal{D}_G^{\text{test}}$ assesses generalization to unseen truss structures under known loading conditions. $\mathcal{D}_{G,L}^{\text{test}}$ tests the model's adaptability to completely unknown conditions, featuring both unseen truss structures and loading scenarios.

**Table B.3**
**Comparison of the normalized root mean square error (NRMSE) evaluated across different datasets with varying energy prediction loss weight** $\lambda_W$. $\mathcal{D}_L^{\text{test}}$ evaluates the model's performance on unseen loading scenarios with known truss structures. $\mathcal{D}_G^{\text{test}}$ assesses generalization to unseen truss structures under known loading conditions. $\mathcal{D}_{G,L}^{\text{test}}$ tests the model's adaptability to completely unknown conditions, featuring both unseen truss structures and loading scenarios.

| NRMSE | $\mathcal{D}_L^{\text{test}}$ | | | $\mathcal{D}_G^{\text{test}}$ | | | $\mathcal{D}_{G,L}^{\text{test}}$ | | |
|---|---|---|---|---|---|---|---|---|---|
| | $S_{11}$ | $S_{22}$ | $S_{33}$ | $S_{11}$ | $S_{22}$ | $S_{33}$ | $S_{11}$ | $S_{22}$ | $S_{33}$ |
| $\lambda_W = 0.0$ | 0.007 | 0.007 | 0.006 | 0.011 | 0.011 | 0.010 | 0.040 | 0.030 | 0.027 |
| $\lambda_W = 0.1$ | 0.006 | 0.005 | 0.006 | 0.010 | 0.009 | 0.010 | **0.034** | **0.026** | 0.025 |
| $\lambda_W = 0.2$ | **0.004** | **0.003** | **0.005** | **0.008** | **0.007** | **0.008** | 0.036 | 0.036 | **0.022** |
| $\lambda_W = 0.5$ | 0.006 | 0.005 | 0.007 | 0.011 | 0.009 | 0.009 | 0.035 | 0.027 | 0.030 |
| $\lambda_W = 1.0$ | 0.005 | 0.005 | 0.006 | 0.010 | 0.009 | 0.009 | 0.036 | 0.028 | 0.026 |

as a regularization term to prevent the model from predicting non-physical stress responses, particularly in unseen loading scenarios, where relying solely on stress data can potentially lead the model to overfit specific patterns, such as abrupt changes or noise in the training data. The value of loss weight $\lambda_W$ balances the emphasis between stress prediction accuracy and the generalization capacity of

**Table C.4**

**Overview of the computational runtime, the software and hardware resources required for different tasks.** The reported runtimes are rough average estimates.

| Tasks | Software | Hardware | Runtime |
| --- | --- | --- | --- |
| Truss dataset generation | Python | CPU (20 cores)[b] | 5 min |
| FE homogenization computations (of the full dataset) | In-house C++ FEM code | CPU (20 cores)[b] | 6 days |
| Model training | PyTorch in Python | GPU[c] | 4.5 h |
| Responses computation using NN model $\mathcal{F}$ [a] | PyTorch in Python | GPU[c] | 0.01 s |
| Responses computation using FE homogenization[a] | In-house C++ FEM code | CPU (10 cores)[b] | 10 min |
| Multiscale simulation with discrete beam elements | In-house C++ FEM code | CPU (10 cores)[b] | 40 min[d] |
| Multiscale simulation with NN-based continuum model | PyTorch in Python | CPU (single core)[b] | 2 min[d] |

[a] Runtimes are reported for a single truss unit cell subject to a defined loading path with 100 load steps.
[b] Computations were performed on the Euler V cluster of ETH Zurich with two 12-core Intel Xeon Gold 5118 processors and 96 GB of DDR4 memory at 2400 MHz.
[c] Computations were performed on a single Nvidia RTX 4090 with CUDA 11.7.
[d] Reported runtimes are averaged for representative examples in the main article only and may vary due to the convergence of the FE solver.

the constitutive model. To illustrate the impact of different $\lambda_W$ values, we present the normalized root mean square error (NRMSE) of the normal components $S_{11}, S_{22}, S_{33}$ of the second Piola–Kirchhoff stress tensor (for simplicity) evaluated on different datasets in Table B.3. It can be observed that choosing a value of $\lambda_W = 0.2$ achieves the best trade-off, as it sufficiently emphasizes the energy term to improve physical plausibility and generalization without significantly sacrificing the model's stress prediction accuracy.

**Appendix C. Computational efficiency**

To demonstrate the efficiency of our approach, we provide the computation runtimes, software, and hardware resources used for the data generation, training, and deployment of our framework in Table C.4. It is important to note that the multiscale simulations with the discrete beam-based structural elements were performed using a highly optimized in-house C++ FEM code [84], whereas the NN-based multiscale simulations were conducted in a pure Python FEM code with a NN as a constitutive model; the latter is inherently slower for large and complex computations [102]. Currently, no general FEM codebase accommodates both discrete beam-based structural simulations and NN-based continuum models, making a direct comparison of computational speeds challenging. Nevertheless, the conceptual advantage of NN-based continuum simulation over a fully resolved discrete simulation in terms of speed is evident from the 20x speedup (as outlined in Table C.4) with the potential for several more orders of magnitude given a competitive implementation that natively integrates FEM and NN-based continuum models.

**Appendix D. Supplementary data**

Supplementary material related to this article can be found online at https://doi.org/10.1016/j.eml.2024.102243.

**Data availability**

The training and test dataset (consisting of truss structures and their effective stress–strain responses) and the pre-trained models are available in the ETHZ Research Collection [103] at https://doi.org/10.3929/ethz-b-000699994.